\documentclass[nohyperref]{article}
\pdfoutput=1
\usepackage{microtype}
\usepackage{graphicx}
\usepackage{booktabs} 

\usepackage{url}
\usepackage{hyperref}

\usepackage[accepted]{icml2022}

\usepackage{amsmath}
\usepackage{amssymb}
\usepackage{mathtools}
\usepackage{amsthm}
\usepackage{multirow}

\usepackage[capitalize,noabbrev]{cleveref}

\usepackage{subfig}
\usepackage{bm}

\theoremstyle{plain}

\theoremstyle{definition}

\theoremstyle{remark}

\usepackage[textsize=tiny]{todonotes}

\begin{document}

\twocolumn[
\icmltitle{Understanding Scaling Laws for Recommendation Models}




\begin{icmlauthorlist}
\icmlauthor{Newsha Ardalani}{M2}
\icmlauthor{Carole-Jean Wu}{M2}
\icmlauthor{Zeliang Chen}{M2}
\icmlauthor{Bhargav Bhushanam}{M2}
\icmlauthor{Adnan Aziz}{M2}
\end{icmlauthorlist}

\icmlaffiliation{M2}{Meta}

\icmlcorrespondingauthor{Newsha Ardalani}{new@fb.com}

\icmlkeywords{Machine Learning, ICML}

\vskip 0.3in
]



\printAffiliationsAndNotice{}  

\begin{abstract}
Scale has been a major driving force in improving machine learning performance, and understanding scaling laws is essential for strategic planning for a sustainable model quality performance growth, long-term resource planning and developing efficient system infrastructures to support large-scale models. In this paper, we study empirical scaling laws for DLRM style recommendation models, in particular Click-Through Rate (CTR).  
We observe that model quality scales with power law plus constant in model size, data size and amount of compute used for training. We characterize scaling efficiency along three different resource dimensions, namely data, parameters and compute by comparing the different scaling schemes along these axes. We show that parameter scaling is out of steam for the model architecture under study, and until a higher-performing model architecture emerges, data scaling is the path forward. The key research questions addressed by this study include: Does a recommendation model scale sustainably as predicted by the scaling laws? Or are we far off from the scaling law predictions? What are the limits of scaling? What are the implications of the scaling laws on long-term hardware/system development? 

\end{abstract}

\begin{figure}[t]
\centering
\includegraphics[width=.48\textwidth]{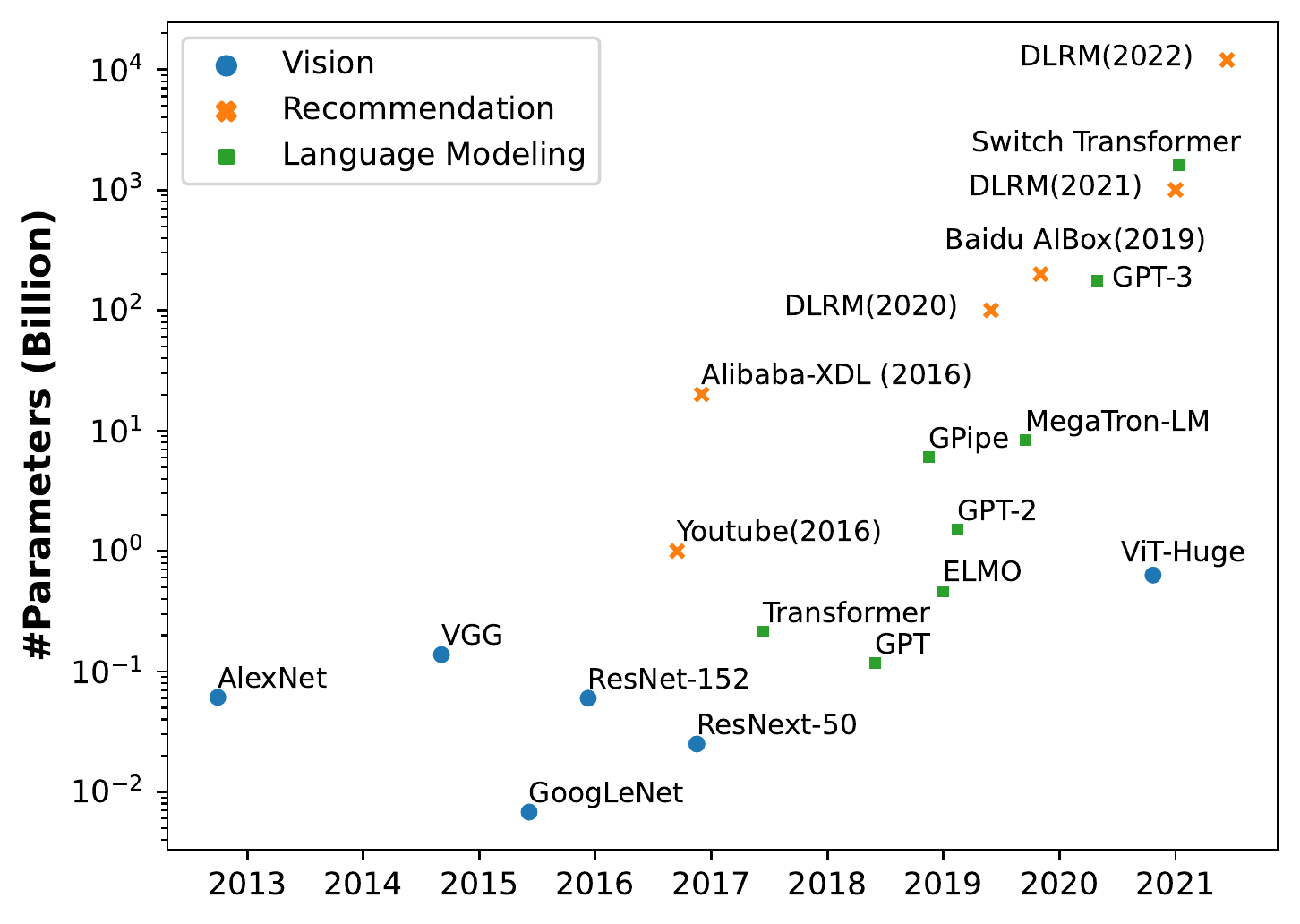}
\caption{Deep learning in general and deep learning based recommendation models in particular have witnessed an exponential growth in parameter size in recent years~\cite{sevillaProgressMachineLearning2021, mudigere2022software, lian2021persia}.
Note the difference in the growth trend across different domains.}
\label{fig:param_growth}
\end{figure}
\vspace{-0.7cm}
\section{Introduction}\label{sec:intro}
Over the last decade, deep learning in general, and deep learning based recommender models (DLRM) in particular, have witnessed an exponential growth in dataset size, model size and system resources~\cite{elkahky2015multi, covington2016deep, sullivan2016faq, liu2017related, yi2018factorized, zhou2019deep, zhao2019aibox, naumov2020deep, zhao2020distributed, lui2021understanding, acun2021understanding, steck2021deep, lian2021persia}, pushing the AI industry into a trillion-parameter era.
Enabling a trillion-parameter model requires a heavy investment in AI systems infrastructure~\cite{mudigere2022software}.
From the system design perspective, the main question/concern is how to scale up, which scaling scheme provides a better return-on-investment (ROI), and how to strategically combine different scaling schemes to provide a better ROI.

Figure~\ref{fig:param_growth} shows a 10000$\times$ growth in model size for language modeling tasks as well as DLRMs over a period of 5 years (2016 - 2021). These results only reflect the growth in published models. We expect that DLRMs have grown even at a faster rate. 
Recommender systems are the major revenue source for many Internet companies. Hence, the details of such models are often confidential.
Recent studies show that just over a period of 2 years (2019 - 2021), recommendation models at Facebook have scaled by $20\times$ in number of parameters, $2.4\times$ in training set size and system infrastructure has grown by 2.5-2.9$\times$~\cite{wu2021sustainable, mudigere2022software}, and more than 50\% of AI training cycles in data-centers are devoted to recommendation models ~\cite{acun2021understanding}.
Despite their importance, there is a limited understanding of how DLRM models scale.
Identifying and understanding a model's scaling property is crucial for designing AI systems and infrastructures that serves such models. Our paper is the first attempt to address this gap.

Recent work~\cite{hestness2017deep, kaplan2020scaling, hernandez2021scaling, henighan2020scaling, gordon2021data, zhai2021scaling, brown2020language, hestness2019beyond, prato2021scaling, bahri2021explaining} shows highly predictable scaling trends in a wide range of domains, including language modeling, machine translation, vision transformers, transfer learning and other autoregressive models. However, it is unknown how recommendation systems scale. Also prior studies exclude embedding parameters in their scaling analysis. Embedding parameters account for a large fraction ($>90\%$) of recommendation model capacity, therefore, it is imperative to study their impact on model quality performance scaling.

Our goal in this work is to characterize scaling laws for deep learning recommendation models, in particular Click-Through Rate (CTR) prediction models. CTR models are some of the most important machine learning tasks in recommender systems, providing personalized experience for billions of users. By studying many different model sizes N (ranging across three orders of magnitude), compute budgets C (ranging across five orders of magnitude), and dataset sizes D (ranging across three orders of magnitude), we demonstrate that a simple \textbf{power law plus constant} explains the relationship between CTR model performance at one epoch and N, D and C.

Figure~\ref{fig:model_arch} shows an overview of a canonical DLRM architecture. At high level, there are two primary components which can be scaled: embedding tables and multi-layer perceptrons (MLPs). Embedding tables can be scaled vertically (increasing the number of embedding rows for each table), or horizontally (widening the dimension of embedding). MLP layers can be scaled by making layers wider or deeper. We study empirical scaling laws for recommendation systems on the normalized cross-entropy loss across four scaling approaches: scaling embedding tables (vertically and horizontally), scaling top MLP layers (which we refer to as overarch layers) as well as scaling all MLP layers (including dense layers, overarch layers, and dense-sparse interaction layers by increasing the width).

\begin{figure*}[tbp]
     \centering
     \subfloat[][\textbf{Overarch Layer Scaling}]{\includegraphics[width=.9\textwidth]{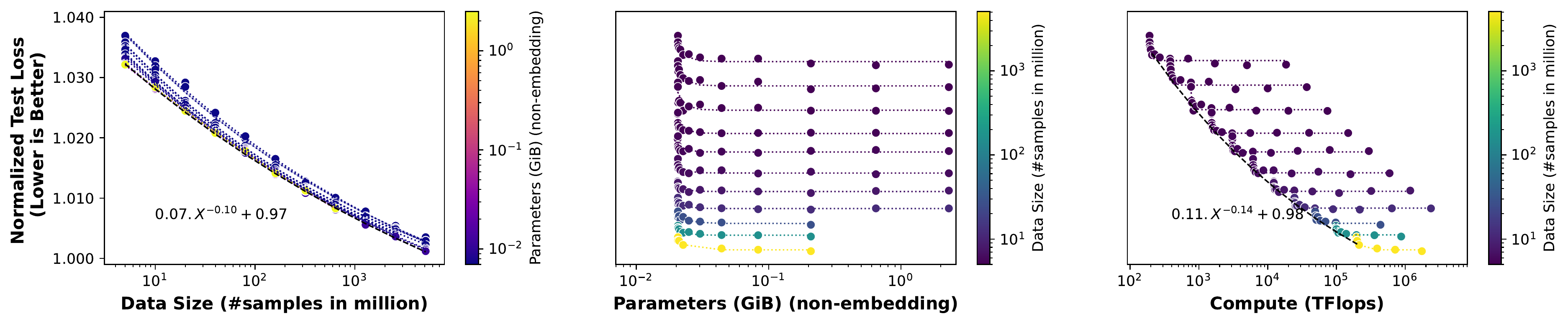}\label{fig:1a}} \\
     \subfloat[][\textbf{MLP Layer Scaling}]{\includegraphics[width=.9\textwidth]{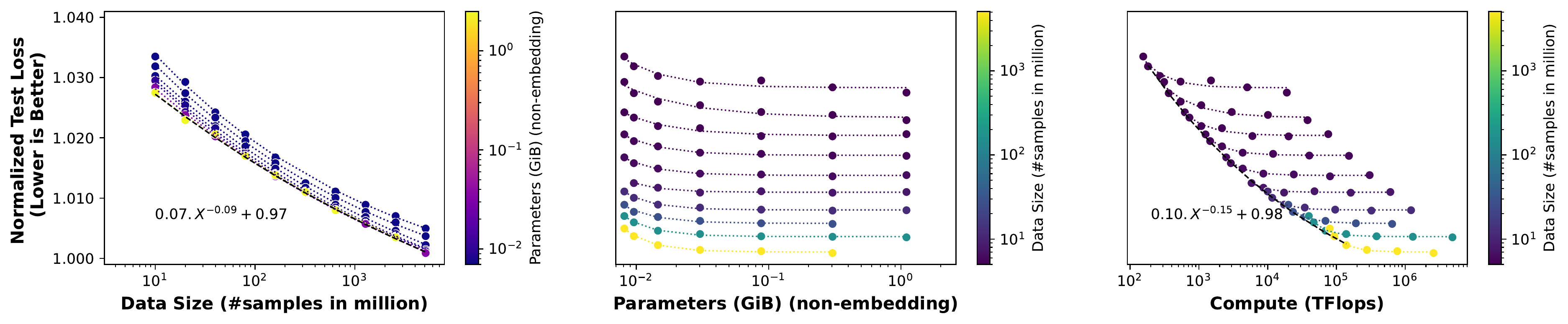}\label{fig:1b}}\\
     \subfloat[][\textbf{Horizontal Embedding Scaling}]{\includegraphics[width=.9\textwidth]{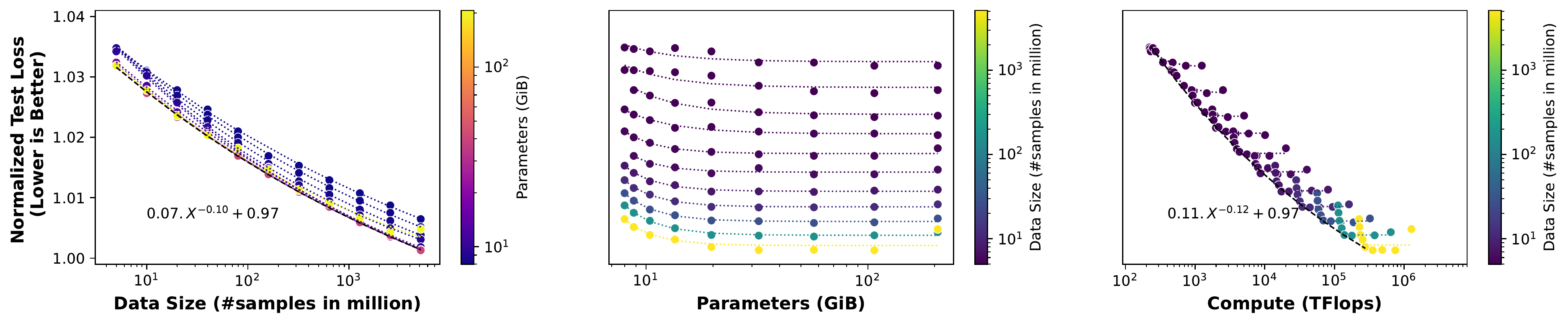}\label{fig:1c}}\\
     \subfloat[][\textbf{Vertical Embedding Scaling}]{\includegraphics[width=.6\textwidth]{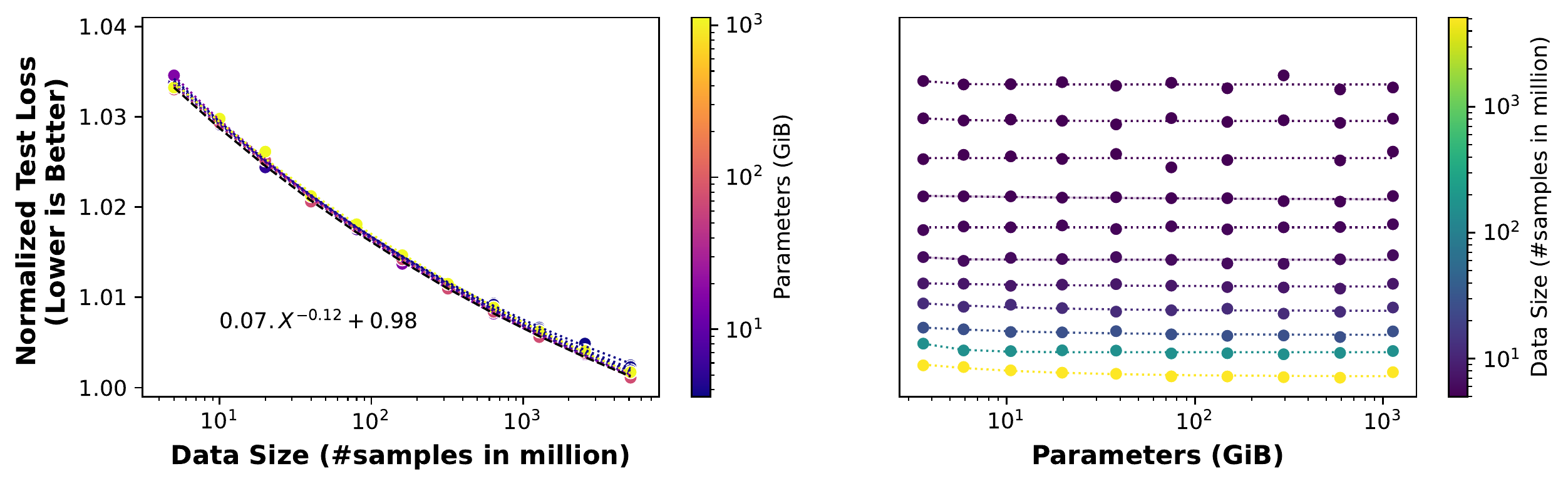}\label{fig:1d}}
\caption{
 Recommendation system’s performance scales with power law plus constant as we increase data size, model size and compute flops for training. 
 \textbf{(a)} scaling model size through increasing MLP layers’ width. \textbf{(b)} scaling model size through increasing overarch layers’ width. \textbf{(c)} scaling model size through increasing embedding table dimensions. \textbf{(d)} scaling model size through increasing the number of rows in embedding tables.
}
\label{fig:intro}
\end{figure*}
\begin{table*}[htbp]
\small
\centering
\begin{tabular}{l|l|l|l|l|l|l|l|}
\cline{2-8}
                                                            & \multicolumn{1}{c|}{$\bm{\alpha}$}  & \multicolumn{1}{c|}{$\boldsymbol{\beta}$}   & \multicolumn{1}{c|}{$\boldsymbol{\gamma}$} & 
                                                            \multicolumn{1}{c|}{{$\bm{R^2}$}} &
                                                            \multicolumn{1}{c|}{{\textbf{Sat.}}} &
                                                            \multicolumn{1}{c|}{\textbf{\begin{tabular}[c]{@{}c@{}}Best Scaling \\ Approach\end{tabular}}}
                                                            & \multicolumn{1}{c|}{\textbf{Ref.}}\\ \hline \hline
\multicolumn{1}{|l|}{\textbf{Data Scaling Efficiency}}      &  0.07 &  [0.09 - 0.12 ] & [0.97 - 0.98] & 0.999 & No & $V > H > O > M$   & Fig~\ref{fig:dse}        \\ \hline
\multicolumn{1}{|l|}{\textbf{Parameter Scaling Efficiency}} & (0 - 0.5]       & [0.4 - 7.6] &  [0.97 - 1] & 0.9 & Yes & $V \approx H \approx O \approx M$ & Fig~\ref{fig:pse}      \\ \hline
\multicolumn{1}{|l|}{\textbf{Compute Scaling Efficiency}}   & 0.11 & {[}0.12-0.15{]} & 0.98 & 0.999 & No & $M > O > H$  & Fig~\ref{fig:cse}       \\ \hline
\end{tabular}
\caption{Scaling efficiency comparison of different scaling techniques: 
Scaling curves are explained by $L(x) = \alpha x^{-\beta} + \gamma$, where x can be dataset size (D), model size (P) or amount of compute flops (C). $\alpha$, $\beta$ and $\gamma$ are parameters of the pareto frontier curve fit. We compare four different scaling schemes; vertical embedding scaling (V), horizontal embedding scaling (H), overarch layers scaling (O) and MLP layers scaling (M) along three efficiency axis. The symbol $>$ signifies better than. $R^2$ shows the goodness of fit. Column 5 (Sat.) shows whether or not the curve is in a saturated regime.}
\label{tab:summary}
\end{table*}
\begin{figure}[htbp]
\centering
\includegraphics[width=.4\textwidth]{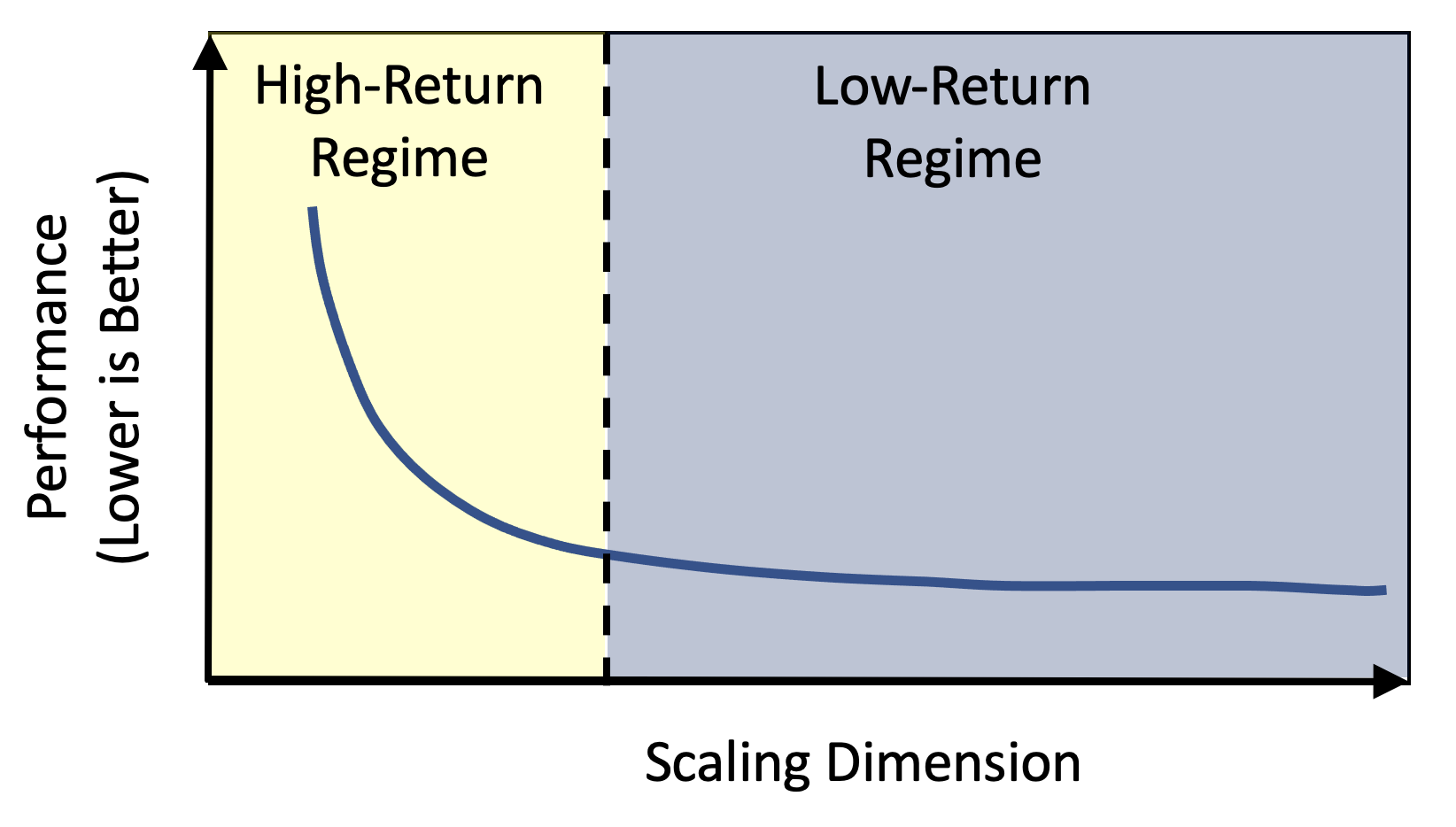}
\caption{Power-Law Function Characterization.}
\label{fig:power_law}
\end{figure}
\subsection{Summary}

Our key findings for CTR prediction models are as follows:

\textbf{Power Law Plus Constant}: We observe that recommendation models performance (test loss) after training for one epoch follows a power law plus constant relationship ($\alpha x^{-\beta} + \gamma$) with resource investment (see Figure~\ref{fig:intro}). Resources include dataset size, model size and the amount of compute flops. The constant $\gamma$ in the power law plus constant function identifies the limits of scaling: that is the best we can achieve if we can hypothetically scale resources to infinity. Table~\ref{tab:summary} shows empirically collected $\alpha$, $\beta$ and $\gamma$ for different scaling schemes and different resource investment scenarios. 

\textbf{Two Phases of Power-law Function}: As illustrated in Figure~\ref{fig:power_law}, power-law functions can be characterized by a high-return phase followed by the slow-return/saturating phase. The point of diminishing return is where the transition happens. If using power-law functions to compare the efficiency of two scaling schemes, one need to pay attention to the exponent of power-law function ($\beta$) as well as the operating phase. Power-law functions with larger magnitude in exponent decay faster and are better for scaling. However, a scaling approach operating within a saturating phase is an inferior technique to a non-saturating approach regardless of its exponent. 

\textbf{Performance Depends Strongly on Dataset Size and Compute and Weakly on Model Parameter Size}: model performance scales strongly with the number of samples in the training set (D) and the amount of compute flops (C), while scales weakly with the number of parameters (P).

\textbf{Limits of Scaling}: The constant ($\gamma$) in the power-law trend captures the irreducible error. This implies that the best normalized test loss to achieve by scaling resources (model parameters, datasize and/or compute flops) to infinity will saturate at 0.98.

\begin{figure*}[t]
\centering
\includegraphics[width=.7\textwidth]{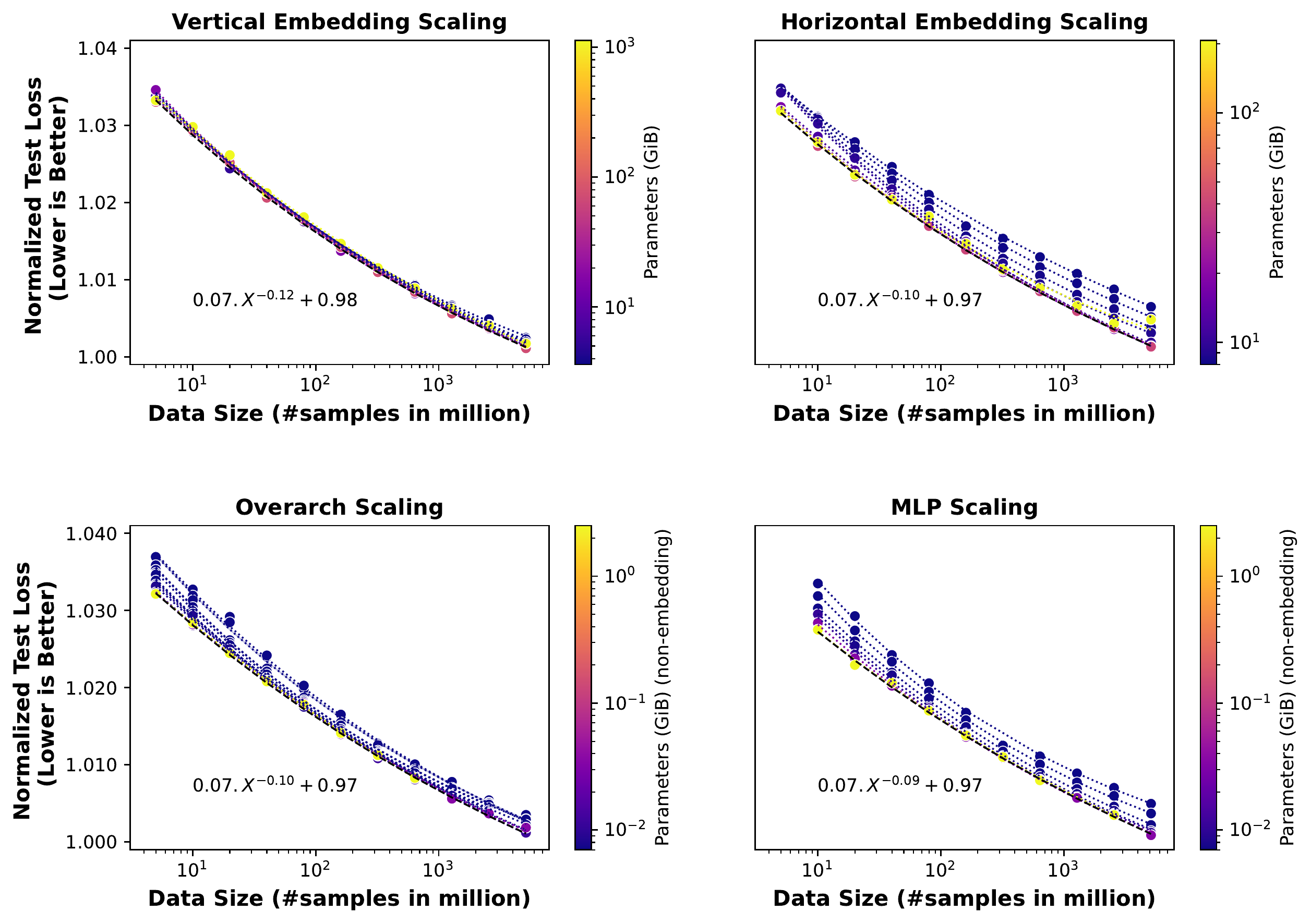}
\caption{Data Scaling Efficiency across different model scaling schemes. While each line shows data scaling trend for a constant model size, the dashed line in each plot along with the equation, captures the pareto optimal curve. As shown, irrespective of the scaling scheme, all models have more or less the same power law scaling profile (power ~ -0.1) when scale model and data together, implying that data scaling efficiency is the same across all model scaling schemes.}
\label{fig:dse}
\end{figure*}

\textbf{Data Scaling Efficiency}: Data scaling efficiency is similar across all scaling schemes ($\beta$ ranging within [0.09, 0.12]) and is insensitive to model size. All scaling schemes operate within the high-return phase. Based on the power law exponents shown in Figure~\ref{fig:dse}, one can see that vertical embedding table scaling (V) is better than horizontal embedding table scaling (H), which itself is better than overarch layer scaling (O), which in turn is better than MLP layer scaling (M), in terms of data scaling efficiency. 
This implies that under a fixed parameter budget, scaling model performance through scaling dataset size and model size in tandem is somewhat sensitive to the parameter scaling approach. 

\textbf{Compute Scaling Efficiency}:  Compute scaling efficiency is similar across all scaling schemes ($\beta$ ranging within [0.12, 0.15]). All scaling schemes operate within the high-return phase. Based on the power-law exponents shown in Figure~\ref{fig:cse}, one can see that MLP scaling is more compute effective than overarch scaling and overarch scaling is slightly more compute effective than embedding dimension scaling.

\textbf{Parameter Scaling Efficiency}: 
Parameter scaling efficiency is different across different scaling schemes ($\alpha$ ranging within [0.4, 7.6]). However, all scaling schemes are operating within the saturating phase (see Figure~\ref{fig:pse}). For an industry-scale model, all parameter scaling techniques are similar in terms of parameter scaling efficiency. 
This implies that under a fixed data budget, scaling model performance through growing the number of parameters in the model is insensitive to the parameter scaling approach.  

\section{Scaling Efficiency}\label{sec:efficiency}
Given a constant budget/resource, the main question is which scaling scheme can provide a better return-on-investment (ROI). We characterize scaling efficiency for three different resources, namely data, parameter, and compute flops. We show that all scaling schemes have similar data scaling and compute scaling efficiency and there are still room for improvement. On the other hand, parameter scaling efficiency is very low as it has already surpassed the point of diminishing return. 
\subsection{Data Scaling Efficiency}
To study data scaling efficiency, we keep model size constant while scaling dataset sizes across a wide range (three orders of magnitude). Conceptually, the slope of the line captures how effectively the model absorbs new information as new data samples are thrown at the problem. Results are shown in Figure~\ref{fig:dse}. Each plot captures a different model scaling scheme (vertical embedding, horizontal embedding, over-arch and MLP scaling).

As it is shown across all scaling strategies, recommendation system performance depends strongly on dataset size and weakly on parameter/model size.  This is counter-intuitive and quite interesting. We continue to see the sizes of embedding tables and the number of embedding tables to grow over the past 5 years. These results imply that industry-scale models are operating within an overfitting regime.

While each line in Figure~\ref{fig:dse} shows data scaling trend for a constant model size, the dashed line in each plot captures the pareto frontier line. As shown, irrespective of the scaling scheme, all models have similar power-law trend. This implies that data scaling efficiency is similar across all model scaling schemes.




\textbf{Summary}
Recommendation system performance depends strongly on data size and weakly on parameter/model size. Comparing this to a large-scale language model ~\cite{hestness2017deep, kaplan2020scaling} where performance scales strongly with model size, recommendation systems are weakly sensitive to model size, and this needs to be taken into account while designing systems for next-generation recommendation systems. Data scaling efficiency is similar across all scaling schemes. This implies that the model under study absorbs information from new data at the same rate, irrespective of the underlying scaling scheme.
Input granularity/vocabulary size does not have any significant impact on the scaling trend.
\begin{figure*}[htbp]
     \centering
     \subfloat[][\textbf{Tandem Compute-Data Scaling}]{\includegraphics[width=.9\textwidth]{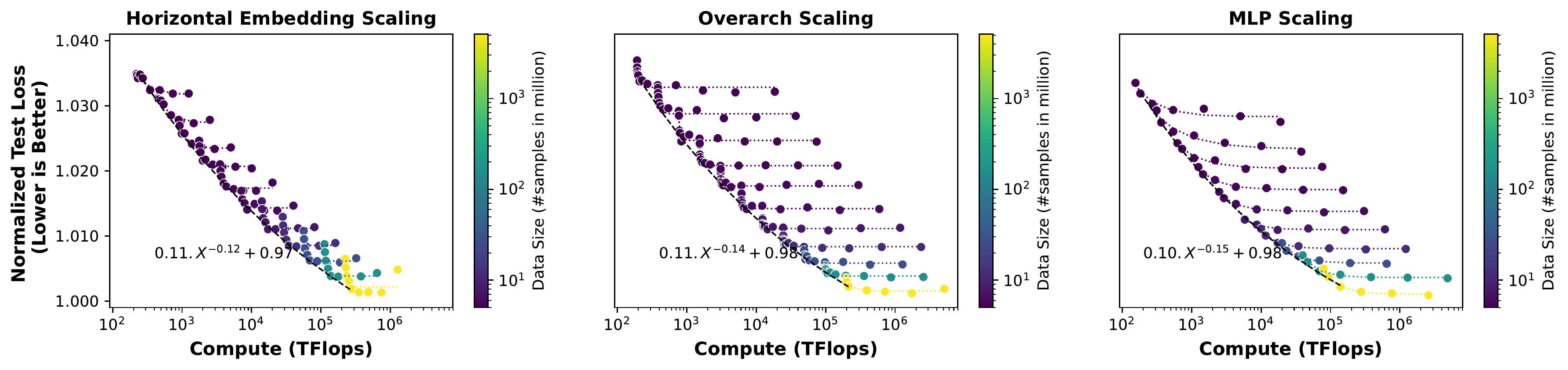}\label{fig:4a}} \\
     \subfloat[][\textbf{Tandem Compute-Parameter Scaling}]{\includegraphics[width=.9\textwidth]{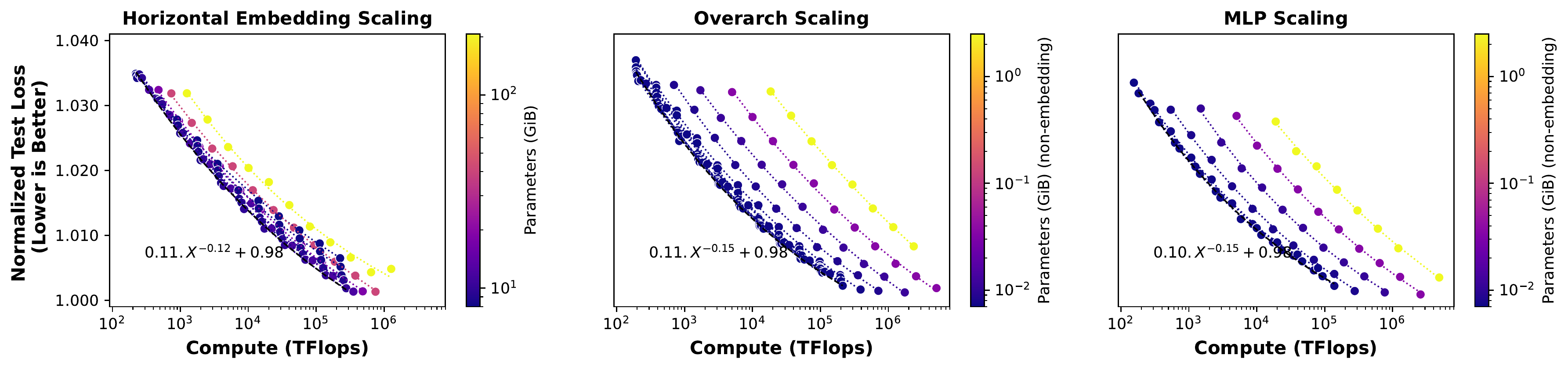}\label{fig:4b}}
\caption{
 Compute Scaling Efficiency -- two views: \textbf{(a)} scaling the amount of compute flops and dataset size in tandem \textbf{(b)} scaling the amount of compute flops and model size in tandem.
}\label{fig:cse}
\end{figure*}
\subsection{Compute Scaling Efficiency}
Our goal is to characterize the slope of the line that captures the relationship between model quality performance and compute flops. Conceptually, the slope of the line captures how quickly the model absorbs new information for the new compute flops thrown at the problem. For compute efficiency analysis, we keep data (or model size) constant, while we scale model size  (or data size). As we scale model size or data size we indirectly increase the amount of compute flops. There is another way to scale compute flops without changing data size or model size and that is to train models for longer. We leave that for future work.

Figure~\ref{fig:cse} shows the results of such scaling. Each plot captures a different model scaling scheme (horizontal embedding, over-arch and MLP scaling. Note that we do not show compute scaling for vertical scaling as increasing the number of rows does not have any impact on compute flops.) As it is shown across all the scaling strategies, recommendation system performance depends strongly on the amount of compute flops.

We present the same results in two different ways: (1) increasing compute flops through model scaling while keeping the data size constant (Figure~\ref{fig:cse}, top row). (2) Alternatively, we increase compute flops through data scaling while keeping the model size constant (Figure~\ref{fig:cse}, bottom row).

\textbf{Scaling Compute and Data in Tandem} 
Figure~\ref{fig:cse}, top row shows the scaling impact of compute flops on performance through scaling model size. Within each line, we keep data size constant while we increase the compute flops through model size scaling.
Note the slight difference in the power of the power law equation across different scaling schemes. It seems that MLP scaling is slightly more effective than overarch scaling, and overarch scaling is slightly more effective than embedding dimension scaling in improving model accuracy for the same amount of increase in the compute budget (0.15 vs. -0.14 vs. -0.12).
Also as shown, at a fixed compute budget, larger dataset sizes results in better performance. Meanwhile, at a fixed accuracy target, smaller dataset sizes are more compute efficient.

\textbf{Scaling Compute and Model Size in Tandem}
Figure~\ref{fig:cse}, bottom row shows the scaling impact of compute flops on performance through scaling data size. Within each line, we keep model size constant while we increase the compute flops through scaling dataset size. As shown, at a fixed compute budget, larger models achieve lower performance. Meanwhile, at a fixed accuracy target, smaller model sizes are more compute efficient.
The dashed line captures the best model size at each compute flop budget that gives the best performance. Figure~\ref{fig:cse} (a) and (b) are basically the same set of points, presented from two different perspective (once grouping points based on dataset size, and once based on model size), therefore, the pareto-optimal line (the dashed line) would be the same.

\textbf{Summary}
At a fixed compute budget, there is a trade-off to be made between spending resources on training models at larger dataset sizes or training models with more parameters. We observe at a fixed compute budget,
models with more parameters show lower/worse performance, and
models trained with larger data sizes show better performance.
From compute efficiency perspective, we observe that at one epoch, MLP scaling is better than overarch scaling, and overrach scaling is better than scaling embedding tables horizontally. Note that scaling embedding tables vertically does not have any impact on compute flops.


\begin{figure*}[htbp]
\centering
\includegraphics[width=.7\textwidth]{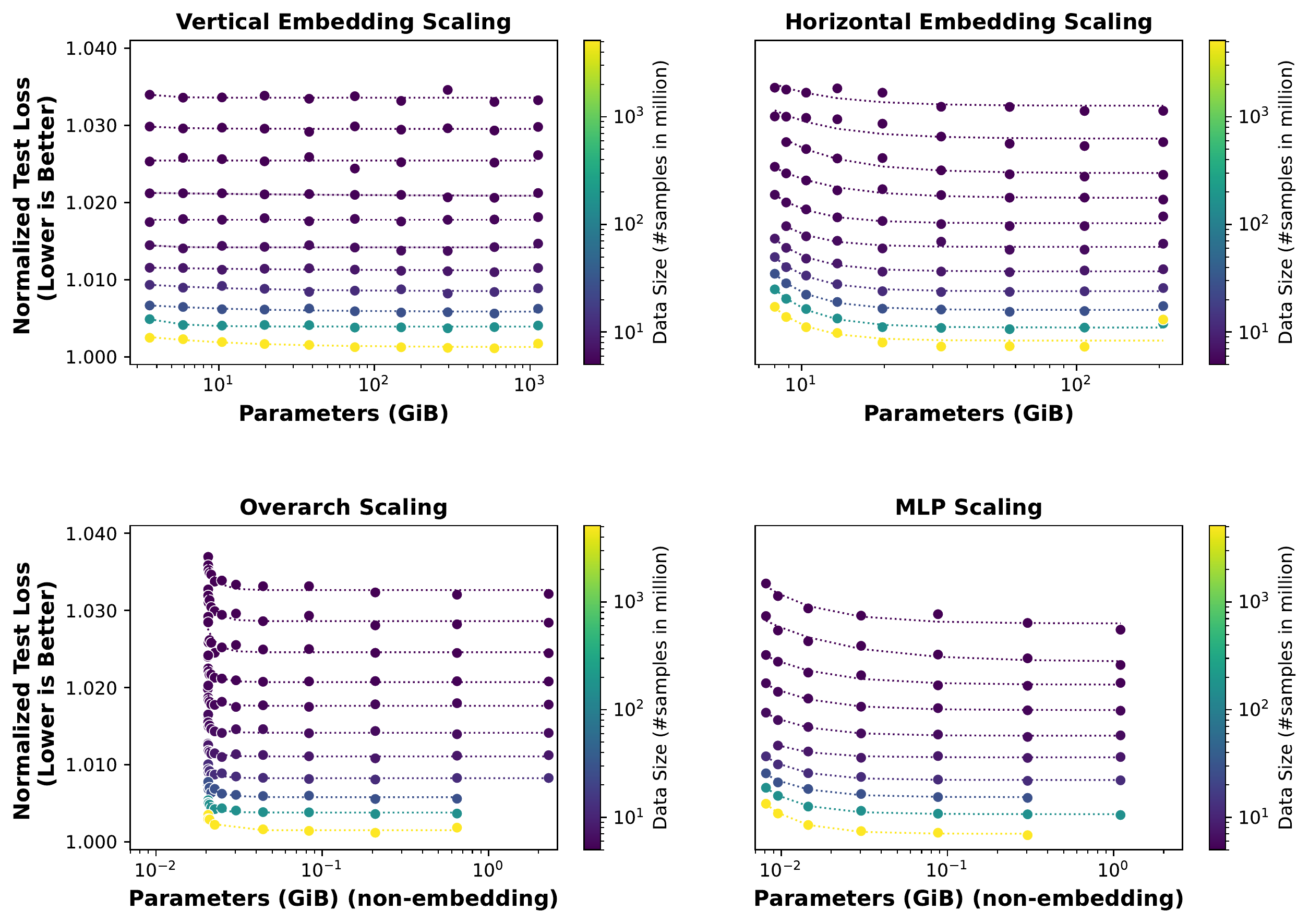}
\caption{Parameter Scaling Efficiency across different parameter scaling schemes. The visible pattern across all scaling schemes is the weak dependence between accuracy and parameter size. 
}
\label{fig:pse}
\end{figure*}
\subsection{Parameter Scaling Efficiency}
Figure~\ref{fig:pse} shows parameter scaling efficiency across different parameter scaling schemes, namely vertical and horizontal embedding scaling, and also over-arch and MLP scaling. MLP scaling involves scaling dense and dense-sparse interaction layers besides overarch layers. We scale MLP layers by increasing the width of each layer.

Parameter efficiency captures how effectively the model performance scales as we increase model capacity/parameter budget. It is widely believed that increasing model capacity would increase model performance. However, much like any power law function this trend will not hold forever and will taper off at some point. Prior work~\cite{kaplan2020scaling, hestness2017deep, hestness2019beyond} shows that language modeling is still at a high-return regime. Unlike language modeling, CTR model capacity has surpassed the point of diminishing return and hence further parameter scaling plays a negligible role in performance improvement.

\begin{figure}[t]
\includegraphics[width=.45\textwidth]{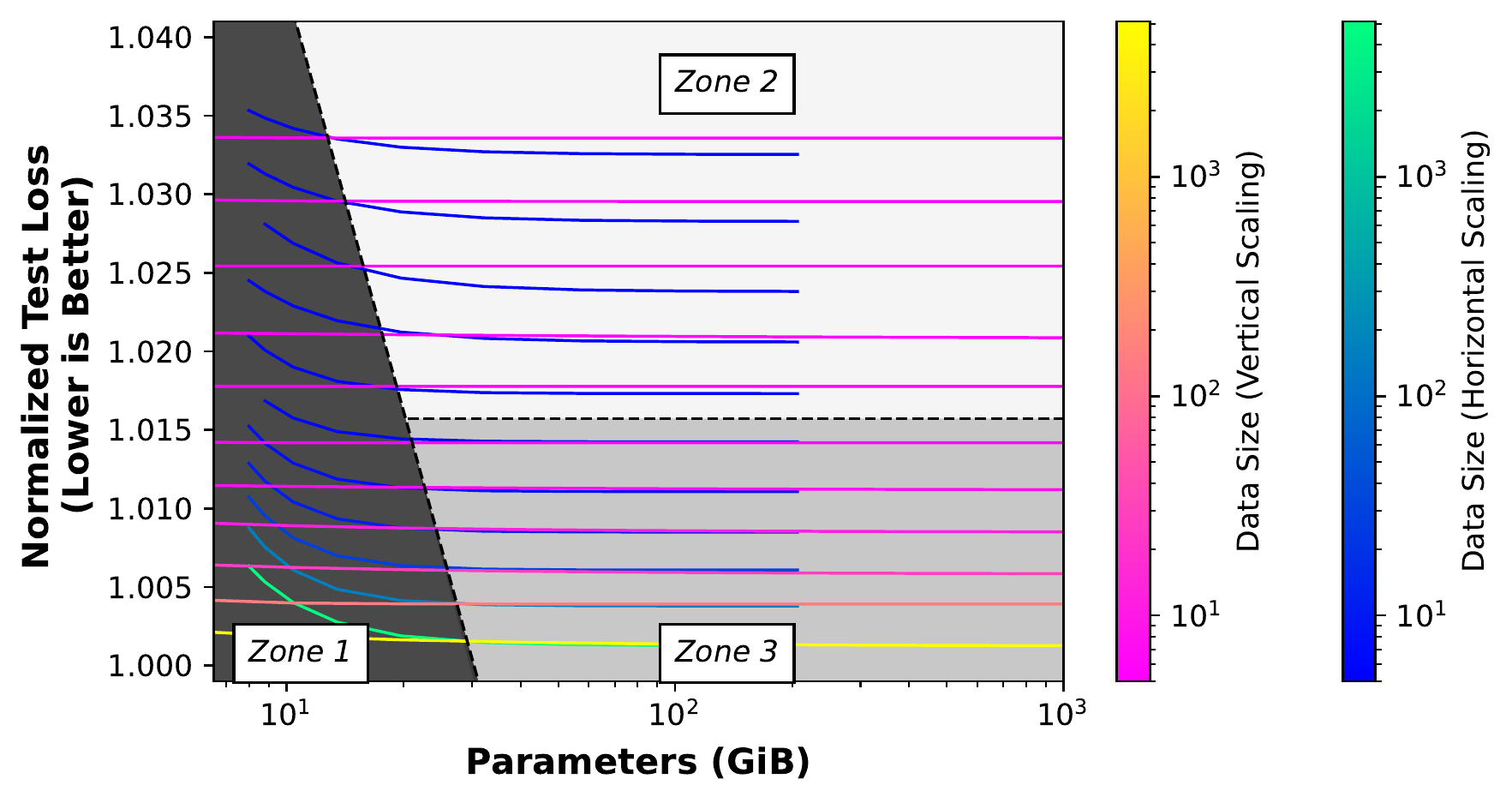}
\caption{When to choose Vertical vs. Horizontal Scaling? 
}
\label{fig:vhs}
\end{figure}
\textbf{Vertical vs. Horizontal Embedding Scaling}
Figure~\ref{fig:vhs} shows the horizontal and vertical scaling results overlaid on the same plot. The main question/concern is for a given parameter budget what is the best scaling strategy. Figure~\ref{fig:vhs} shows that the answer depends on data budget and parameter budget. If the parameter budget is small relative to the data budget, vertical scaling is strongly better than horizontal scaling (zone 1). If the parameter budget is large relative to the data budget, horizontal scaling is better than vertical scaling (zone 2). If we have sufficiently large data and parameter budget, it would be effectively the same to scale models through vertical scaling or horizontal scaling (zone 3).

\begin{figure}[t]
\includegraphics[width=.45\textwidth]{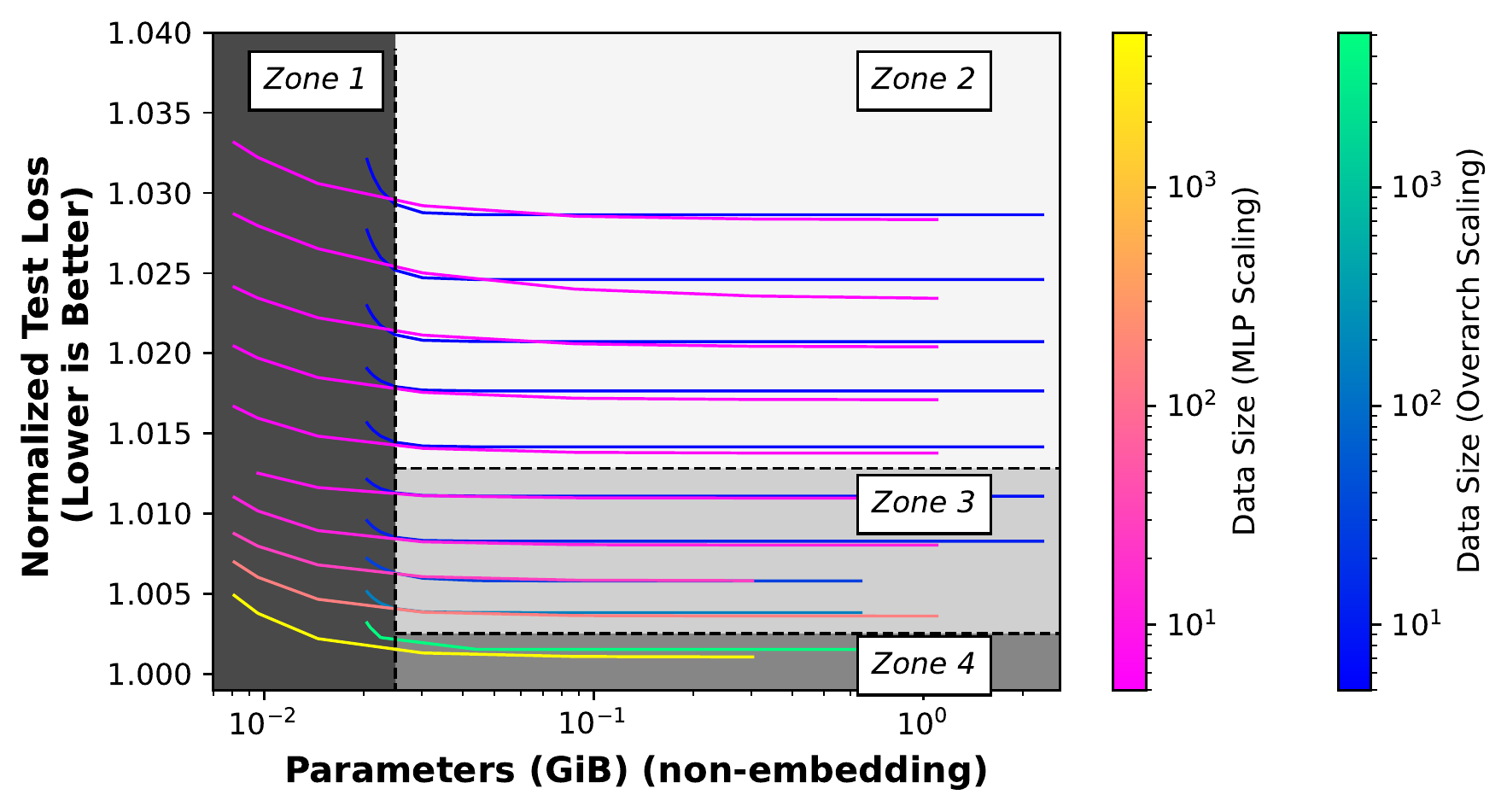}
\caption{When to choose Over-arch vs. MLP Scaling? 
}
\label{fig:oms}
\end{figure}
\textbf{Overarch vs. MLP Layer Scaling}
Figure~\ref{fig:oms} shows the overarch scaling and MLP scaling results overlaid on the same plot. The main question/concern is for a given non-embedding parameter budget what is the best scaling strategy. As shown, the answer depends on data and parameter budget. If the non-embedding parameter budget is small (zone 1) or data budget is small relative to parameter budget (zone 2), MLP scaling is better than overarch scaling (zone 1). For relatively large data and parameter budget (zone 3), it would be effectively the same to scale non-embedding parameters through over-arch scaling or MLP scaling (zone 3). For really large non-embedding parameters budget, MLP scaling is strongly better than overarch scaling (zone 4).

\begin{table}[t]
\small
\def\arraystretch{1.4}
\centering
\begin{tabular}{@{}llllll@{}}
                                                      &                            & \multicolumn{2}{c}{\textbf{Data Budget}}                                                                                                                                                                     &  &  \\
\multicolumn{1}{l}{\parbox[t]{2mm}{\multirow{3}{*}{\rotatebox[origin=l]{90}{\textbf{Param. Budget~~~~~}}}}}                                                       &                            & Small                                                                                            & Large                                                                                            &  &  \\ \cline{3-4}
& \multicolumn{1}{l|}{Small} & \multicolumn{1}{l|}{\begin{tabular}[c]{@{}l@{}}V \textgreater H\\ M \textgreater O\end{tabular}} & \multicolumn{1}{l|}{\begin{tabular}[c]{@{}l@{}}V \textgreater H\\ M \textgreater O\end{tabular}} &  &  \\ \cline{3-4}
\multicolumn{1}{c}{}                                  & \multicolumn{1}{l|}{Large} & \multicolumn{1}{l|}{\begin{tabular}[c]{@{}l@{}}H \textgreater V\\ M \textgreater O\end{tabular}} & \multicolumn{1}{l|}{\begin{tabular}[c]{@{}l@{}}V $\approx$H\\ M $\approx$O\end{tabular}}               &  &  \\ \cline{3-4}
\end{tabular}
\caption{Parameter Scaling Scheme Comparison. V: Vertical embedding table scaling, H: Horizontal embedding table scaling, M: MLP layer scaling, O: Overarch scaling}
\label{tab:param_comp}
\end{table}
\textbf{Summary}
Unlike prior analysis in NLP domain, model capacity scaling plays a negligible role in DLRM performance improvement (for the model architecture under study).
Parameter scaling has been a great scaling scheme in the past, however industry-scale models are outrageously large, operating within the saturated regime.
Table~\ref{tab:param_comp} summarizes our findings across the entire spectrum (small vs. large dataset regime, and also small vs. large parameter budget regime). The symbol “$>$” signifies “better than”. The best approach depends on the operating regime. Production-scale models are outrageously large in model size and dataset size, hence all parameter scaling techniques are similar.


\section{Sensitivity Analysis}\label{sec:sens}
\begin{figure}[t]
\includegraphics[width=.45\textwidth]{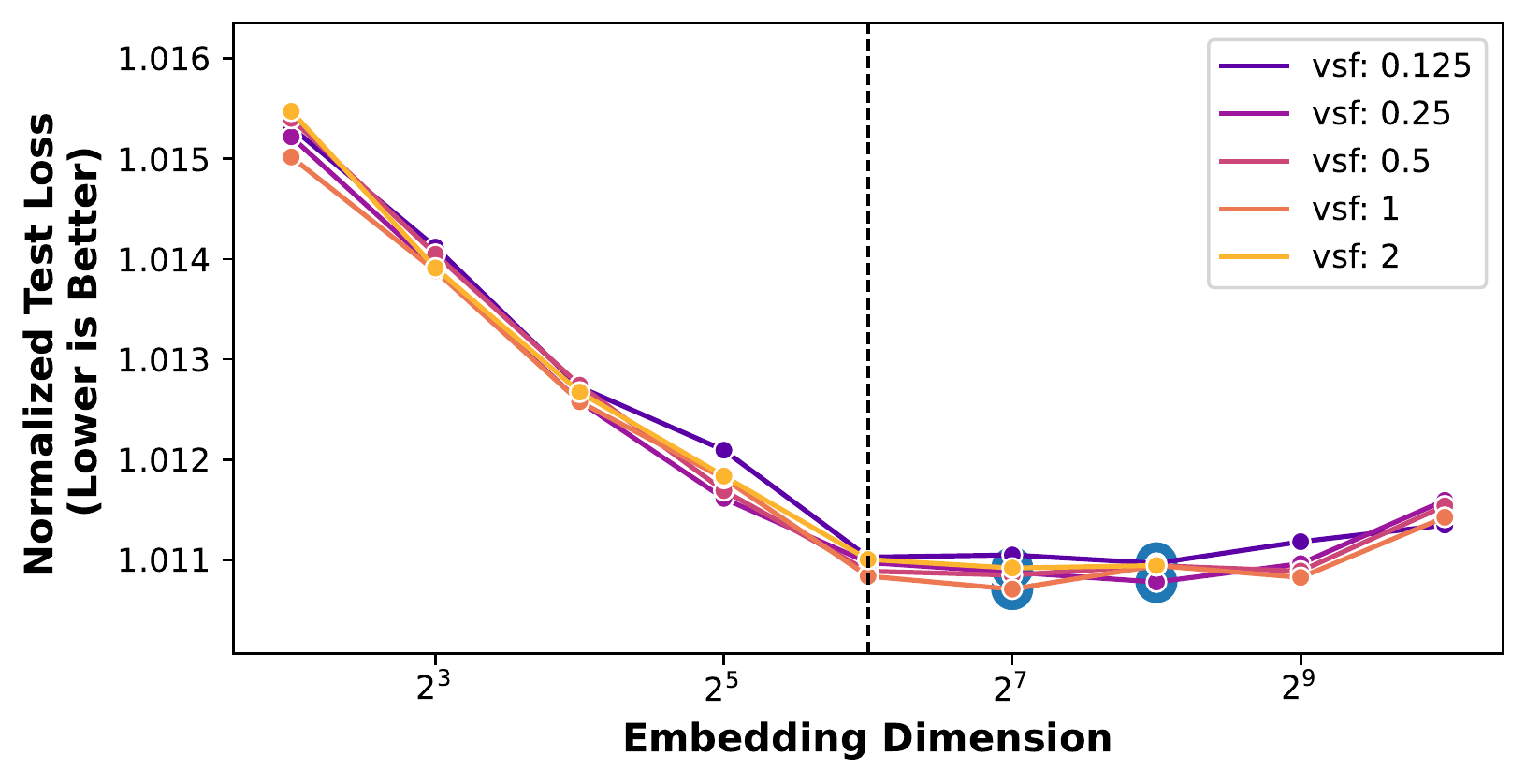}
\caption{Embedding Dimension Sensitivity to Table Size: Each line shows a different vertical scaling factor (vsf).
The large blue circles show the minimum loss for each line. However,  the knee in the curve happens consistently at 64.}
\label{fig:emb_dim}
\end{figure}
\subsection{How to effectively scale embedding dimension with the number of rows?}
Figure~\ref{fig:emb_dim} shows how the best embedding dimension varies as we increase the number of rows in the table (increasing the vertical scaling factor). As shown, the best embedding dimension tends to get smaller as the vertical scaling factor gets larger (256 is the best embedding dimension for vertical scaling factors of 0.125$\times$ and 0.25$\times$ vs. 128 for 0.5$\times$, 1$\times$, and 2$\times$ vertical scaling factors). However, the best performance and the most resource-efficient embedding dimensions are not necessarily the same thing. As shown, the knee in the curve (the point of diminishing return) starts setting off around embedding dimension = 64 for all table sizes. This implies that the resource-efficient design point for embedding dimension has a weak dependency on vertical scaling factor. This result implies going beyond 64 will not offer high ROI from the perspective of resource efficiency.

\begin{figure}[t]
\includegraphics[width=.49\textwidth]{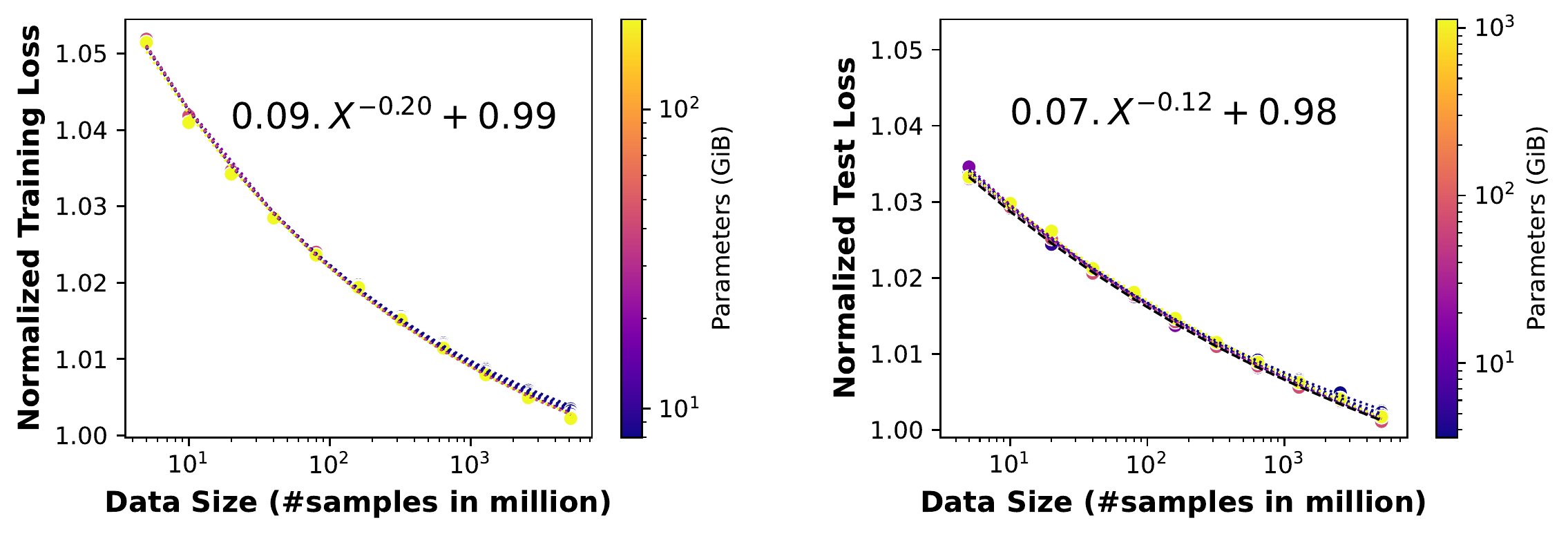}
\caption{Data Scaling Efficiency on Training Loss vs Test Loss. 
Note the difference in power-law exponent between training and test curve.}
\label{fig:tvt}
\end{figure}
\vspace{-0.1cm}
\subsection{Training vs. Test}
As shown in Figure~\ref{fig:tvt}, the learning curve for training data is steeper (-0.20 vs -0.12)  than the learning curve for test data. Both curves capture scaling of the same model trained on the same data but evaluated on two different dataset. The curve on the left is evaluated on data points from the training set, and the model on the right is evaluated on the test set. This gap implies that information absorbed by the model from extra training points is more effective at predicting data from the same distribution (training distribution as opposed to test distribution), which is intuitively expected.

\section{Discussion}\label{sec:disc}
Power-law curves characterizing different scaling schemes offer insights about data efficiency, parameter efficiency and compute efficiency of each scaling technique. One can potentially compare the efficiency of any pairs of scaling techniques (e.g. vertical embedding scaling vs. horizontal embedding scaling) by comparing their power-law curves along three different axis (data, compute, parameter). 
Table~\ref{tab:param_comp} shows the result of such comparisons. 
As depicted, there is no single scaling technique that stands out along all scaling efficiency dimensions. 
For example, horizontal embedding scaling (H) is better than MLP scaling (M) in terms of data efficiency, but is worse in terms of compute efficiency. 


Recent analysis show that just over a period of 5 years, industry-scale recommendation models have grown by four orders of magnitude~\cite{mudigere2022software,lian2021persia}.
Power-law analysis supports this trend in the past. Parameter scaling has the largest magnitude in the exponent when approximated by power-law trend. However, industry-scale recommendation models are far too large and saturated, hence further parameter growth will not offer high ROI from the perspective of resource efficiencies.

Meanwhile data scaling and compute scaling are still within the high diminishing return regimes. This implies that data scaling should be treated as a first-class scaling approach until a better model architecture emerges. That said, we should be mindful that data scaling is not a sustainable approach in the long run (in its raw form) due to limitations on data retention.

To overcome this, we need to think about alternatives. Here are some suggestions some of which we plan to explore as next step: 
(1) Log more data, particularly through logging more negative examples and reducing positive down sampling 
(2) Explore training models using historical data as teacher models to synthesize valuable information learned from the historical data for the use of more recent models. 
(3) Scale data volume horizontally rather than vertically, i.e. adding more features rather than adding more rows. 

Scaling laws can also be used to guide long-term hardware development. 
Hardware design usually starts 3-5 years in advance, relying on an accurate projection of models growth over the next 3-5 years. 
Our analysis suggests that going forward hardware does not need to grow to support larger models. Rather, we need to design hardware/systems to support training with larger dataset sizes. 

Another key take-away is that the constants in the power-law plus constant equations are bounded at 0.98 (loss measured in normalized entropy). This constant captures the model's accuracy at an infinite scaling limit, which can be used as a guideline to gauge how far off the industry-scale models are from the infinite limit. Prior analysis in NLP domain has shown that innovations in model architectures (e.g. transitioning from LSTM to Transformer) can improve the coefficient of the power-law (i.e. $\alpha$ in $\alpha.x^{-\beta} + \gamma$) and shift the curves downward, however they have negligible impact on the exponent ($\beta$) of the power law ~\cite{hestness2017deep, brown2020language}. This suggests that model architecture exploration is a short-term solution for performance growth.
A long-term solution would require improving the exponent of the power-law trend. To this day, it is still an open research question what controls the slope of the power-law. It appears that the slope of power-law curve is unique to each domain irrespective of the model architecture~\cite{hestness2017deep, hestness2019beyond}. Prior analysis suggests that improving data distribution can improve the exponent of power-law~\cite{bahri2021explaining}. 
Recent work suggest that through effective data pruning we can beat power law and achieve exponential scaling~\cite{sorscher2022beyond}.
\section{Conclusion}
This paper is the first effort to explore the scaling properties of recommendation models from a 
holistic perspective, characterizing the scaling efficiency along three different axis (data, compute, parameters) and four different scaling schemes (embedding table scaling vertically and horizontally, MLP layer scaling and over-arch layer scaling). We characterize power-law scaling laws by examining an industry-scale recommendation model. We show that unlike NLP domain, parameter scaling is running out of steam and does not contribute much to performance improvement and until a higher-performing model architecture emerges, data scaling is the path forward. We share the limits of scaling and chart out important directions across all dimensions of data, model and system design.

\bibliography{ref}
\bibliographystyle{icml2022}

\newpage
\appendix
\onecolumn
\begin{figure}[b]
\centering
\includegraphics[width=.5\textwidth]{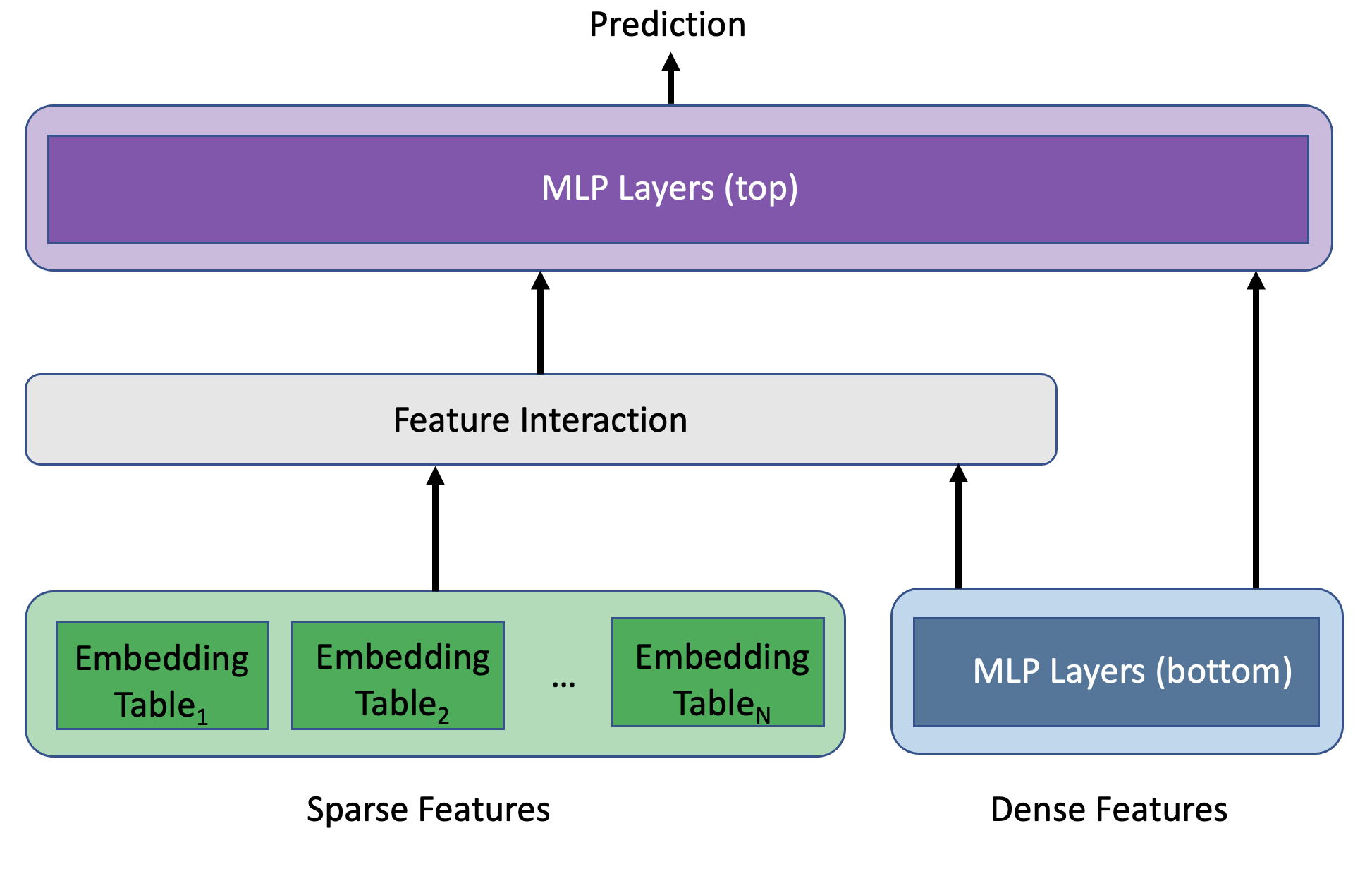}
\caption{A schematic of deep learning model architecture.}
\label{fig:model_arch}
\end{figure}
\section{Methodology}\label{sec:method}
To investigate the predictability of recommendation system performance as parameters/data/amount of compute increase, we train models of various sizes on randomly selected subsets of data. 
We use production-level model and production-level dataset for all analysis. In this section, we provide details on model architecture, input dataset and scaling approaches.

\subsection{Model  Architecture}
The model is similar to open-source DLRM model~\cite{naumov2019deep} in architecture. Figure~\ref{fig:model_arch} gives an overview of a canonical Deep Learning Recommendation Model (DLRM) architecture. DLRMs process user-content pairs to predict the probability that a user will interact with a particular piece of content, commonly referred to as the click-through-rate (CTR). To produce such a prediction, DLRMs consume two types of features: dense and sparse. Dense features represent continuous data, such as a user’s age or the time of day, while sparse features represent categorical data, such as domain names or recent web pages viewed by a user. To encode this categorical data, sparse features are represented as one-hot or multi-hot binary vectors which are only activated for a small subset of relevant categories (hence the term sparse). 

\subsection{Dataset}
The dataset includes timestamp, user specific and recommendation specific features. 
Prior works on industry-scale recommendation models~\cite{zhou2019deep, acun2021understanding, adnan2021accelerating, zhao2022understanding} have shown distinct feature distribution characteristics as compared to open-source CTR data sets~\cite{criteo-1tb, criteo-kaggle}.

\subsection{Performance Metric}
Similar to prior work on industry-scale recommendation models~\cite{he2014practical}, we evaluate the model quality in terms of normalized cross-entropy loss (NE for short).
NE is equivalent to the average log loss per impression
divided by the average true background click through rate
(CTR) for every impression. The lower the NE loss, the better is the prediction
made by the model. Dividing by the entropy of the background CTR makes the NE insensitive to
the background CTR. 
We report normalized loss in all Figures. The loss is normalized to minimum observable NE in all of our experiments.

\subsection{Data Scaling}
We explore datasets of size 5M, 10M, 20M, 40M, 80M, 160M, 320M, 160M, 320M, 640M, 1280M, 2560M and 5120M samples for training. 
The number of datapoints are selected within this range to keep exploration time under a day on 8-16 V100 GPU cards. 
We train all models to one epoch.

\subsection{Parameter Scaling}
We study parameter scaling across four different scaling schemes: Scaling embedding tables vertically, i.e. ding more rows to the tables, scaling embedding tables horizontally, i.e. increasing the embedding dimension, scaling overarch layers’ width, and finally scaling all MLP layers’s width which includes dense layer, overarch layer and dense-sparse interaction layer. 

In the baseline model, different tables have different numbers of rows. Not all tables have the same embedding dimension, and not all layers have the same width. 

For vertical scaling, we scale the number of rows within each table by 1/64x, 1/32x, 1/16x, 1/8x, 1/4x, 1/2x, 1x, 2x, 4x, 8x, while keeping all other parameters constant as the baseline model.

For horizontal scaling, we look at embedding dimensions of various sizes: 4, 8, 16, 32, 64, 128, 256, 512 and 1024. 

For overarch scaling, we scale the overarch layers’ width by 1/256x, 1/128x, 1/64x, 1/32x, 1/16x, 1/8x, 1/4x, 1/2x, 1x, 2x, 4x, 8x, 16x, 32x. 

For MLP scaling, we scale MLP layers’ width by 1/8x, 1/4x, 1/2x, 1x, 2x, 4x, and 8x. Different layers have different width to begin with but they all scale proportionally. 
Since model size is mostly dominated by embedding tables, we exclude embedding parameters in the parameter count for MLP scaling and over-arch scaling analysis. 

For all experiments, We use internal tools to collect the number of parameters. 

\subsection{Compute Scaling}
The amount of compute flops can be scaled by increasing the number of epochs, the number of data-points per epoch or increasing the model size. We train all models for one 
epoch. We scale amount of compute flops by increasing dataset size and/or model size.
For each experiment, we use internal tools to collect the number of compute flops per example and multiply by the total number of examples to get the total number of compute flops.


\end{document}